\documentclass[aps,prb,showpacs,reprint,amsmath,amssymb,superscriptaddress,a4paper]{revtex4-1}
\usepackage[utf8]{inputenc}
\usepackage{amsmath}
\usepackage{amsfonts}
\usepackage{amssymb}
\usepackage{graphicx}

\begin{document}

\title{Epitaxial Pb on InAs nanowires}

\author{Thomas Kanne}
\email{thomas.kanne@nbi.ku.dk}
\affiliation{Center For Quantum Devices, Niels Bohr Institute, University of Copenhagen, 2100 Copenhagen, Denmark}

\author{Mikelis Marnauza}
\affiliation{Center For Quantum Devices, Niels Bohr Institute, University of Copenhagen, 2100 Copenhagen, Denmark}

\author{Dags Olsteins}
\affiliation{Center For Quantum Devices, Niels Bohr Institute, University of Copenhagen, 2100 Copenhagen, Denmark}

\author{Damon J. Carrad}
\affiliation{Center For Quantum Devices, Niels Bohr Institute, University of Copenhagen, 2100 Copenhagen, Denmark}

\author{Joachim E. Sestoft}
\affiliation{Center For Quantum Devices, Niels Bohr Institute, University of Copenhagen, 2100 Copenhagen, Denmark}

\author{Joeri de Bruijckere}

\affiliation{Center For Quantum Devices, Niels Bohr Institute, University of Copenhagen, 2100 Copenhagen, Denmark}
\affiliation{Kavli Institute of Nanoscience, Delft University of Technology, 2628 CJ Delft, The Netherlands}

\author{Lunjie Zeng}
\affiliation{Department of Physics, Chalmers University of Technology, Gothenburg, Sweden}

\author{Erik Johnson}
\affiliation{Center For Quantum Devices, Niels Bohr Institute, University of Copenhagen, 2100 Copenhagen, Denmark}

\author{Eva Olsson}
\affiliation{Department of Physics, Chalmers University of Technology, Gothenburg, Sweden}

\author{Kasper Grove-Rasmussen}
\affiliation{Center For Quantum Devices, Niels Bohr Institute, University of Copenhagen, 2100 Copenhagen, Denmark}

\author{Jesper Nyg\aa rd}
\email{nygard@nbi.ku.dk}
\affiliation{Center For Quantum Devices, Niels Bohr Institute, University of Copenhagen, 2100 Copenhagen, Denmark}

\begin{abstract}

Semiconductor-superconductor hybrids are widely used for realising complex quantum phenomena such as topological superconductivity and spins coupled to Cooper pairs. Accessing exotic regimes at high magnetic fields and increasing operating temperatures beyond the state-of-the-art requires new, epitaxially matched semiconductor-superconductor materials. The challenge is to generate favourable conditions for heterostructure formation between materials with the desired inherent properties. Here, we harness increased knowledge of metal-on-semiconductor growth to develop InAs nanowires with epitaxially matched, single crystal, atomically flat Pb films along the entire nanowire. These highly ordered heterostructures have a critical temperature of 7~K and a superconducting gap of 1.25~meV, which remains hard at $8.5$~T, thereby more than doubling the available parameter space. Additionally, InAs/Pb `island’ devices exhibit magnetic field-driven transitions from Cooper pair to single electron charging; a pre-requisite for use in topological quantum computation. Introducing semiconductor-Pb hybrids potentially enables access to entirely new regimes for an array of quantum systems.
\end{abstract}
\maketitle

The development of high-quality semiconductor-superconductor heterostructures underlies the pursuit of new types of quantum bits based on exotic physics arising at the hybrid interface \cite{lutchyn2018majorana, krogstrup2015epitaxy, mourik2012signatures, DasNatPhys12, Deng2012NL, deng2016majorana, ZhangNature18, PendharkarTinArxiv, KlinovajaPRB2014}. Aside from the ability to host novel two-level systems \cite{LarsenPRL15, LuthiPRL18, TosiPRX2019, HaysPRL2019, PradaarXiv2019}, when the underlying materials exhibit strong spin-orbit coupling, a high $g$-factor and `hard gap' induced superconductivity\cite{chang2015hard}, they are ideally suited for supporting topological superconductivity under applied magnetic field\cite{OregPRL10, LutchynPRL10,  mourik2012signatures, deng2016majorana, albrecht2016exponential, DasNatPhys12, Deng2012NL, lutchyn2018majorana,PradaarXiv2019}. Signatures of topological superconductivity in these systems have been studied almost exclusively using InAs or InSb as the semiconductor and Al and Nb(TiN) as the superconductor, with the large spin-orbit coupling and $g$-factor ensuring the suitability of these semiconductors\cite{lutchyn2018majorana}. By contrast, there is strong interest in improving the superconductor material to broaden the scope of hybrid applications\cite{CarradArXiv191100460Cond-Mat2019, bjergfelt2019superconducting, PendharkarTinArxiv}, since Al has a relatively low critical temperature, $T_\mathrm{C}$ and field $B_\mathrm{C}$ \cite{chang2015hard}, while the soft gap of Nb-based hybrids has
thus far prevented isolated hybrid segments supporting Cooper pair charging \cite{albrecht2016exponential, CarradArXiv191100460Cond-Mat2019, shen2018parity, PendharkarTinArxiv, Vaitiekenasfullshell18}; a crucial requirement for topologically protected qubit schemes\cite{AasenPRX16}. The main goal is therefore development of epitaxial hybrids exhibiting a large, hard gap resilient to high magnetic fields and elevated temperatures\cite{CarradArXiv191100460Cond-Mat2019, bjergfelt2019superconducting, PendharkarTinArxiv}.

\begin{figure*}
  \centering
    \includegraphics[width=17cm]{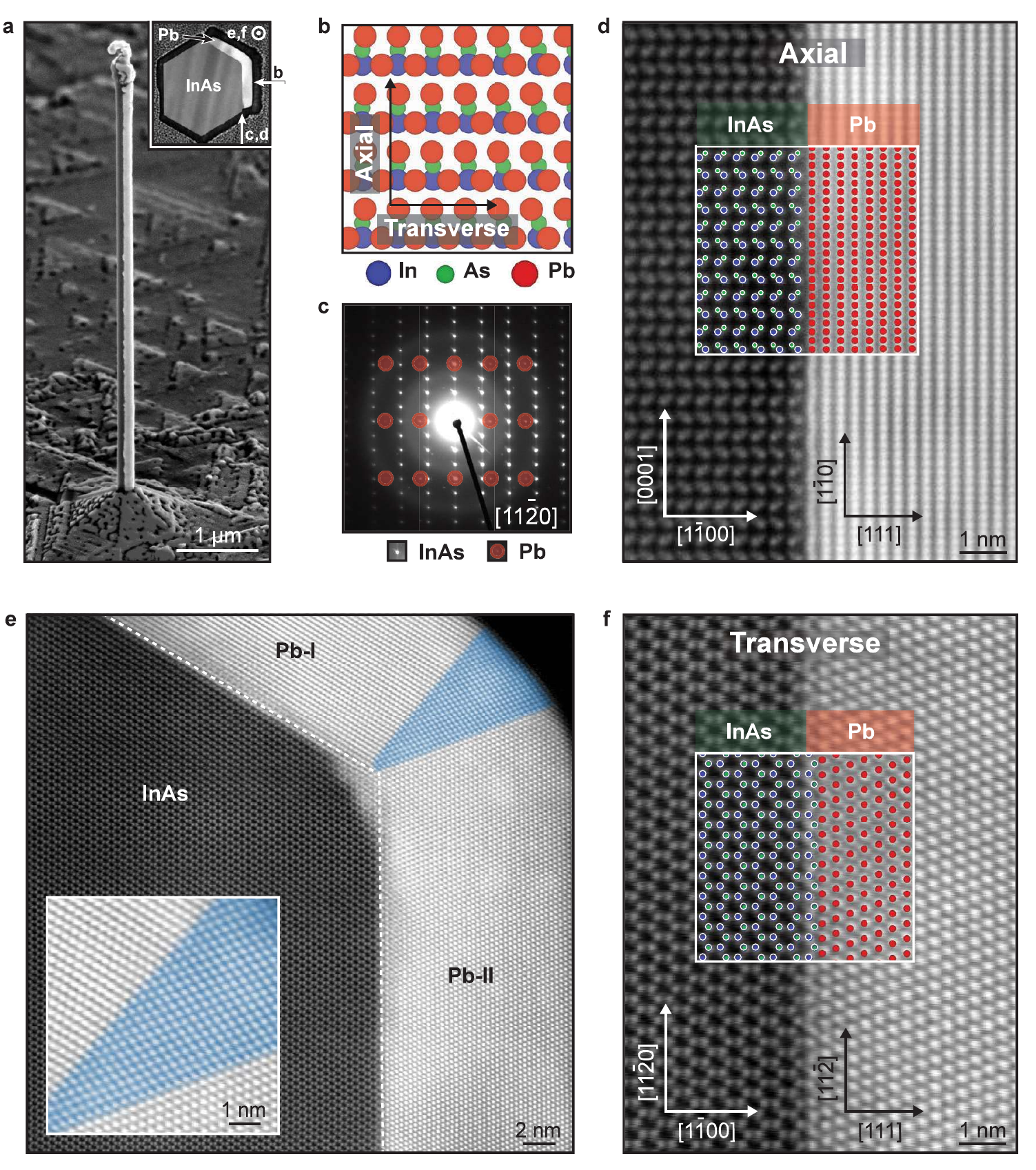}
    \caption{\textbf{InAs/Pb bi-crystal interfacial match.} \textbf{a}, Scanning electron micrograph of an InAs nanowire with a 10~nm Pb thin film on two facets. Inset shows a HAADF STEM micrograph of a cross-sectioned InAs/Pb/Si heterostructure. \textbf{b}, Simulated relaxed bi-crystal match between InAs (blue, green) and Pb (red) viewed normal to the nanowire facet. \textbf{c}, Selected area electron diffraction pattern of an entire nanowire along the $[11\Bar{2}0]/[11\Bar{2}]$ direction. Pb diffraction peaks are marked with red semi-transparent circles.  \textbf{d}, Atomic resolution HAADF STEM micrograph of the InAs/Pb interface along the $[11\Bar{2}0]/[11\Bar{2}]$ direction parallel to the nanowire facet. Simulated relaxed bi-crystal interfacial match overlayed and marked with a white box. \textbf{e}, HAADF STEM micrograph along the $[0001]/[1\Bar{1}0]$ direction of the corner between two adjacent facets, highlighted with white dotted lines. Two single Pb crystals are connected by a wedge shaped single crystal (false coloured blue). \textbf{f}, Atomic resolution HAADF STEM micrograph of the InAs/Pb interface viewed along the transverse direction as in \textbf{e}. Theoretically relaxed bulk bi-crystal interfacial structure superimposed on interface.} 
    \label{fig1}
  \hfill
\end{figure*}

\begin{figure*}
  \centering
    \includegraphics[width=17cm]{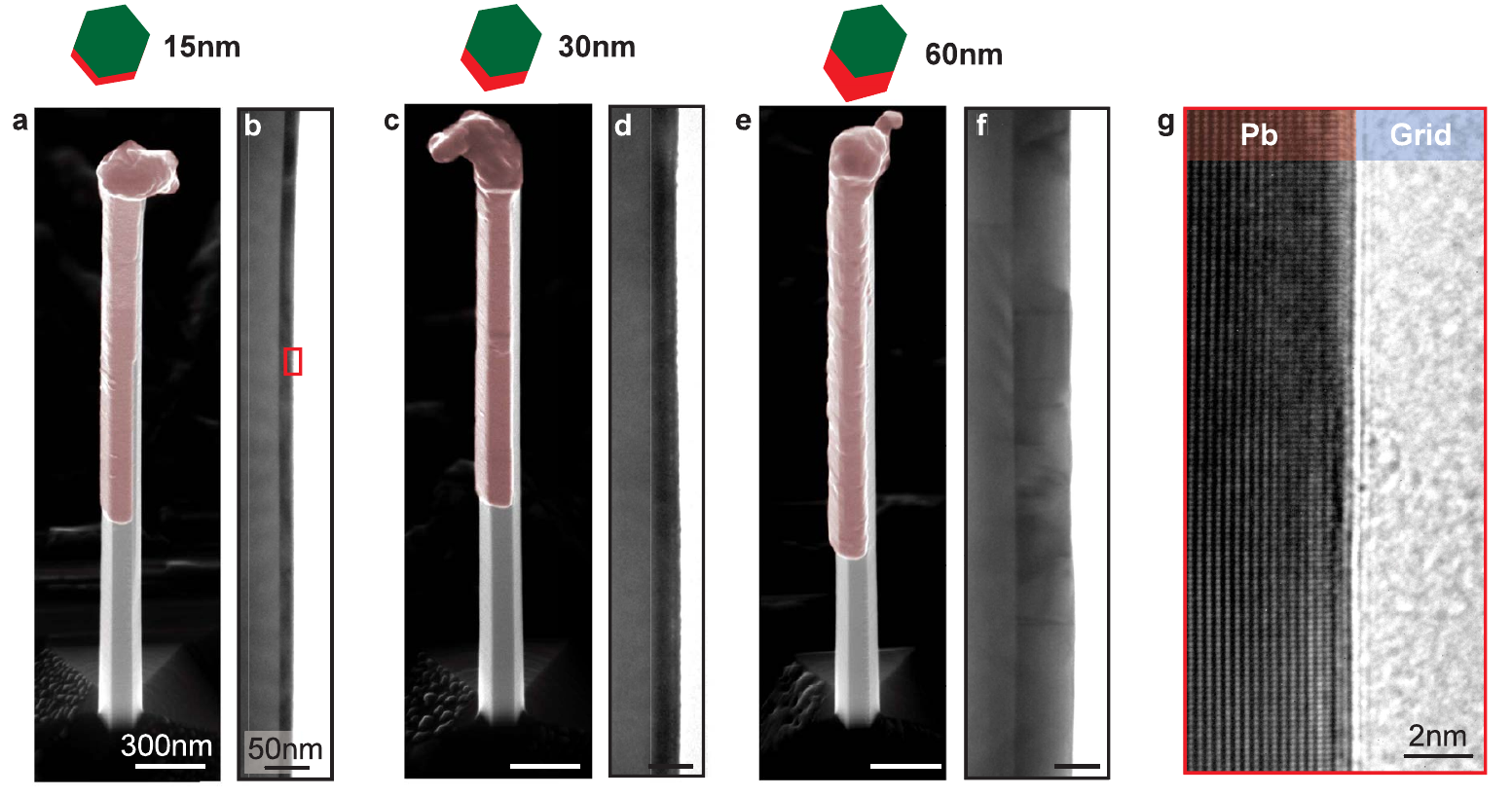}
    \caption{\textbf{Pb thickness dependent morphology.} Pb grown on two facets of the InAs nanowires masked by adjacent nanowires. \textbf{a},\textbf{c} and \textbf{e} show surface sensitive SEM micrographs of 15~nm, 30~nm and 60~nm in-situ deposited Pb films, respectively. The micrographs have the same scale shown in \textbf{a}. \textbf{b},\textbf{d} and \textbf{f} present LR-TEM micrographs of same thicknesses and same scale shown in \textbf{b}. \textbf{g} HR-TEM micrograph of the terminating layer of a 15~nm thin film marked with a red square in \textbf{b}.} 
    \label{fig2}
  \hfill
\end{figure*}

In this article we present hybrid nanowires utilising Pb as the superconductor, which possesses the highest bulk $T_\mathrm{C}$ and $B_\mathrm{C}$ of all elemental type-I superconductors, vastly extending the hybrid parameter space. We deposit the Pb films without breaking vacuum after InAs nanowire growth\cite{krogstrup2015epitaxy,  GuskinNanoscale17, CarradArXiv191100460Cond-Mat2019, bjergfelt2019superconducting}, and ensure carefully controlled conditions which promote strong thermodynamic driving forces towards single crystal formation. From transmission electron microscopy (TEM) we find a low energy domain match, as expected from modelling of the InAs/Pb interface, which enables the growth of an atomically flat single Pb crystal along the entire nanowire facet. All previous semiconductor-superconductor nanowire heterostructures exhibited axial grain boundaries and/or polycrystallinity\cite{krogstrup2015epitaxy, GuskinNanoscale17, CarradArXiv191100460Cond-Mat2019, bjergfelt2019superconducting, PendharkarTinArxiv}. 
The single crystal Pb suppresses the potential for impurity-generated discontinuity of the hybrid wavefunction\cite{deng2016majorana}, which may occur in granular films\cite{GuskinNanoscale17, CarradArXiv191100460Cond-Mat2019, bjergfelt2019superconducting, PendharkarTinArxiv}. Further, the theoretical description used to optimise the epitaxial structures through consideration of thermodynamic driving forces and structural factors leads to increased understanding of metal/semiconductor epitaxy. The epitaxial InAs/Pb structure yields a hard induced superconducting energy gap of $\Delta \sim 1.25$~meV, corresponding $T_\mathrm{C} \sim 7$~K, and $B_\mathrm{C}$ exceeding $8.5$~T, the highest reported values for epitaxial semiconductor-superconductor hybrids. In addition, InAs/Pb `island' devices support Cooper pair charging due to the hard gap. At finite magnetic field, this evolves into single electron transport through a hybrid bound state\cite{albrecht2016exponential, shen2018parity, CarradArXiv191100460Cond-Mat2019, Vaitiekenasfullshell18, sestoft2018engineering, PendharkarTinArxiv}; a crucial feature in the context of topological quantum computation. The epitaxial structure and dramatic enhancement of the accessible parameter space represented by InAs/Pb hybrids make them prime candidates for use in a wide range of semiconductor-superconductor hybrids based on nanowires\cite{mourik2012signatures,deng2016majorana, LarsenPRL15, TosiPRX2019, LuthiPRL18}, selective area grown networks\cite{VaitiekenasPRL18sag, AseevNL18}, and planar heterostructures\cite{KjaergaardNatComm16, ShabaniPRB2016}. Further, Pb possesses features that distinguish it from other elemental superconductors, such as two-band, strong-coupling superconductivity and a large spin-orbit coupling, which may lead to the emergence of new phenomena in semiconductor-superconductor physics \cite{lutchyn2018majorana, PradaarXiv2019, SauNatComm2012,SuNatCommun2017,KlinovajaPRB2014, LarsenPRL15, TosiPRX2019}. 

\section*{InAs/Pb epitaxy}
Aside from the attractive physical properties, the choice of Pb was based on analytical considerations of the central stages in metal film growth on different semiconductor nanowires. To do this, we theoretically assessed structural stability in combination with estimates of the interplay between the four contributions to the overall excess energy, which determine the thermodynamic driving forces and kinetic limiting factors. A full description is given in Supplementary Section 1 and an additional 12 elemental superconductors on InAs nanowires are theoretically studied in Supplementary Section 2. For low temperature depositions ($T<150$~K) it is energetically favourable for Pb to grow in a single crystallographic orientation with respect to InAs. Combined with the high grain boundary mobility of Pb, this suggests that growth of Pb on InAs would result in large crystals with an epitaxial match to the nanowire. In contrast, epitaxial Al grows in 2 or 4 orientations on InAs depending on thickness\cite{krogstrup2015epitaxy}, and other superconductor materials employed so far feature no long-range order\cite{CarradArXiv191100460Cond-Mat2019, bjergfelt2019superconducting, GuskinNanoscale17, PendharkarTinArxiv}, consistent with our findings in Supplementary Section 2. These findings suggest Pb as the optimal elemental superconductor material to combine with InAs. Although we focus on InAs/Pb nanowires in this article, we note that Pb will likely match well to other semiconductors (see Supplementary Section 3.1), and our results should translate to planar heterostructures\cite{ShabaniPRB2016,KjaergaardNatComm16} and selective area grown networks\cite{ VaitiekenasPRL18sag, AseevNL18}. Further, the framework presented in Supplementary Section 1 is instructive in developing the range of semiconductor-metal epitaxial structures. 

Figure \ref{fig1}a shows a scanning electron microscope (SEM) micrograph of a hexagonal InAs nanowire with Pb deposited on two facets. The wurtzite InAs nanowires were grown along the $\{ 0001 \}$ direction using molecular beam epitaxy (MBE), with flat $\{1\Bar{1} 00\}$ facets. The hexagonal morphology was optimised as outlined in Methods and Supplementary Section 3. After the nanowires were grown the wafer was transferred without breaking ultra-high vacuum to a second chamber where the substrate was placed on a liquid nitrogen cooled substrate holder, $T \sim 120$K, for several hours, before Pb was deposited using e-beam evaporation\cite{CarradArXiv191100460Cond-Mat2019, bjergfelt2019superconducting, GuskinNanoscale17}. The inset in Figure \ref{fig1}a shows a low magnification high angle annular dark field scanning transmission electron microscope (HAADF STEM) micrograph of a cross-sectioned InAs/Pb/Si heterostructure. Figure \ref{fig1}a shows the modeled relaxed hetero-epitaxial match along $\{ 1\Bar{1}00 \}$(InAs)/$\{ 111 \}$ (Pb) directions and the arrows indicate the axial and transverse direction. Figure \ref{fig1}c shows a selected area electron diffraction (SAED) pattern of an entire nanowire, oriented along the $\{ 11 \Bar{2}0 \}$/$\{11\Bar{2} \} $ direction parallel to the nanowire facet. Notably, only one Pb orientation relative to the InAs nanowire facet is observed (red circles), confirming the theoretical hypothesis that a Pb thin film forms a single crystal along an entire nanowire facet. 

Figure \ref{fig1}d shows a high resolution (HR) HAADF STEM micrograph of the InAs/Pb interface where the orientation is similar to that of Fig.~\ref{fig1}c. The inset in Fig.~\ref{fig1}d shows how the theoretically predicted strain relaxed  hetero-epitaxial match fits to the observed interface. From this, a particularly small bi-crystal interfacial domain is found along the $\{ 0001 \}$/$\{ 1\Bar{1}0 \}$ nanowire growth direction, with two planes of Pb for each plane of InAs and a bulk residual mismatch of only $0.35\%$. From HR TEM investigations of the entire nanowire length no edge-dislocations were found, indicating that the small strain needed to obtain the epitaxial domain is absorbed. Figure \ref{fig1}e presents a HAADF STEM micrograph of the cross-sectioned InAs/Pb nanowire shown in Fig. \ref{fig1}a (inset). The two grains ($I$, $II$) form single crystals along their entire nanowire facet, and are merged by a wedge-shaped single-crystalline domain, false coloured blue. The inset in Figure \ref{fig1}e shows how the wedge-shaped grain accommodates two coherent grain boundaries and thus reduces the grain boundary energy between the single $I$ and $II$ crystals. Strain in InAs and Pb along the transverse direction is observed by comparing the STEM micrograph and the predicted strain relaxed hetero-epitaxial match in Figure \ref{fig1}f. By utilizing the average [$1\Bar{1}00$] plane spacing as a scale to measure the relative stress in the structure, Pb along the [$11\Bar{2}$] direction was measured to be compressively strained $\sim$ 1.7 \% whereas InAs along [$11\Bar{2}0$] direction was found to be tensile strained $\sim$ 0.7 \%. Assuming that the InAs [$1\Bar{1}00$] planes are not heavily influenced by the interface, the compressed Pb and expanded InAs indicate that the transverse bi-crystal interfacial match seeks a domain of two interface planes of InAs for three planes of Pb. More detail is provided in Supplementary Section 3. The presented single crystal nature and epitaxial relation relative to the InAs nanowire facets was found for all investigated Pb thicknesses (5-60 nm) and Pb thus appears robust against stress driven terms such as incoherent grain boundaries.

To study the Pb film morphology, surface sensitive scanning electron microscopy (SEM) micrographs of InAs/Pb hybrids are shown in Figs~\ref{fig2}a,c,e for 15, 30 and 60~nm Pb films, respectively. Patterning of the Pb coatings was implemented to enable comparative analysis of core/shell morphology and achieved by adjacent nanowires acting as shadow masks during Pb deposition\cite{krizek2017growth, gazibegovic2017epitaxy, PendharkarTinArxiv}. Figure~\ref{fig2}b,d,f show bright field TEM micrographs for the different Pb thicknesses, and Fig.~\ref{fig2}g shows a HR TEM micrograph of the terminating planes of a 15 nm film of Pb marked with a red square in Fig.~\ref{fig2}b. The Pb morphology is continuous and atomically flat for most of the investigated film thicknesses (10 nm - 30 nm). However, above a critical thickness (in our case $\sim$ 50 nm), dependent on the growth conditions, the surface becomes more rough (Fig.~\ref{fig2}e). In Fig.~\ref{fig2}f, bending contours and contrast changes are observed, indicative of strain fields in the Pb film. This may be caused by the increased strain imposed by the grain boundaries between the adjacent facets or a slight bending of the nanowires, likely due to either thermal expansion coefficient differences or axial strain \cite{krogstrup2015epitaxy, bjergfelt2019superconducting}. 

Obtaining a single crystal superconductor on a single crystal semiconductor is critical for hybrid and topological applications to ensure that each end of e.g. the topological system coincides with the geometrically defined end of the superconducting segment, rather than at an impurity\cite{deng2016majorana}. Achieving the structures presented here required detailed understanding of how to control and promote epitaxial growth, via thermodynamic and grain growth kinetics considerations. We focus specifically on Pb, with more general explanations provided in Supplementary Section 1-3. The aim during the initial phase of Pb thin film growth is to promote layer-by-layer-like growth. For island growth this is achieved by decreasing the critical cluster size and thereby increasing the initial concentration of critical clusters \cite{pentcheva2003non}. Experimentally, this was done by lowering the substrate temperature and increasing the material flux. After the initial clusters have been established they start to grow. In this phase, interface and strain energy may result in recrystallization of the clusters in a kinetically limited process due to the thermodynamic driving forces. In a later stage the clusters start to impinge and coalesce into a single crystal, which is now strongly dependent on the main contributions to the overall excess energy \cite{venables1983nucleation, vesselinov2016crystal}. We find that it is especially important to have very well defined nanowire facets to decrease the interface and strain energy (Supplementary Section 1.1). The micrographs in Figs~\ref{fig2}a,c suggest that after a continuous film of Pb has grown, the growth switches to layer-by-layer mode. Here, small islands on the completed atomically flat layer act as nucleation sites for the next layer. Upon warming to room temperature and subsequent exposure to oxygen there is a chance that the Pb thin film dewets since it grows far from thermodynamic equilibrium \cite{thompson2012solid}. We find that the film morphology is improved when the substrate remains on the cold holder for approximately 10 hours after terminating growth. Dewetting and oxidiation of Pb thin films may also be prevented by growing a capping layer with a lower dewetting probability. 

\begin{figure}
  \centering
    \includegraphics[scale=1.05]{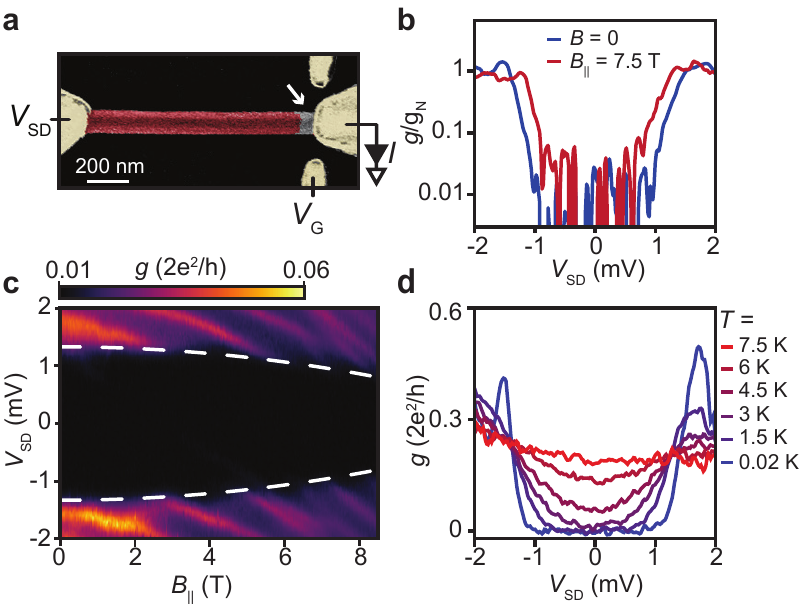}
    \caption{\textbf{Hybrid InAs/Pb tunnel spectroscopy.} \textbf{a}, False coloured SEM micrograph showing the InAs/Pb hybrid nanowire tunnel spectroscopy device with Ti/Au contacts and gates (yellow), Pb (red) and bare InAs nanowire segment (grey; also indicated by white arrow). The measurement circuit consists of source-drain bias $V_\mathrm{SD}$, measured current $I$, and the gate voltage $V_\mathrm{G}$, used to control the tunnel barrier. \textbf{b}, Differential conductance, $g$, normalised to the normal state conductance, $g_\mathrm{N} = g|_{V_\mathrm{SD} = 2\mathrm{mV}}$, as a function of $V_\mathrm{SD}$, showing 100-fold conductance suppression within $\pm\ \Delta\ \sim\ 1.25$~meV at $B = 0$ (blue) and $\pm\ \Delta\ \sim\ 1.1$~meV for $B_{\parallel} = 7.5$~T (red). \textbf{c}, $g$ as a function of $V_\mathrm{SD}$ and parallel magnetic field, $B_{\parallel}$, showing the persistence of highly suppressed conductance until 8.5~T. Dashed white lines serve as guide to the eye. \textbf{d}, $g$ as a function of $V_\mathrm{SD}$ recorded at increasing $T$ from 0.02 to 7.5 K. %A fit to the data (indicated by dashed white curves) gives critical field $B_{\parallel,\mathrm{C}}$ $\sim$ 13~T. The evolution of the superconducting gap as a function of temperature shows a $T$\textsubscript{C} $\sim$ 7~K.
    } 
    \label{trans1}
  \hfill
\end{figure}

\section*{Tunneling spectroscopy of InAs/Pb junctions}
Having obtained the highly ordered epitaxial InAs/Pb heterostructures, we expect hard gap superconductivity with the inherited properties of the Pb film\cite{chang2015hard}. To characterise this we performed electron transport measurements on normal metal-superconductor (NS) tunnel junctions with the device geometry shown in Fig.~\ref{trans1}a. Prior to normal metal contact deposition, the Pb was selectively etched using H$_2$O, such that a bare InAs segment remained between the normal metal and superconductor (Methods and Supplementary Section 5). This segment acts as a tunnel barrier, tunable using side-gate voltage $V_{\mathrm{G}}$\cite{chang2015hard,mourik2012signatures}. Figure \ref{trans1}b shows differential conductance, $g$, normalised to the normal state conductance, $g_\mathrm{N} = g|_{V_\mathrm{SD} = 2\mathrm{mV}}$, as a function of source-drain bias voltage $V_{\mathrm{SD}}$ at zero magnetic field (red) and finite parallel magnetic field, $B_{\parallel} = 7.5$~T (blue). The strong $g$ suppression observed for $|V_\mathrm{SD}|<1.25$~mV reflects the superconducting energy gap, $\Delta$, in the single-particle density of states of the InAs/Pb nanowire. The 100-fold suppression is similar to the hard superconducting gaps reported in III/V nanowires with epitaxial Al\cite{chang2015hard, gazibegovic2017epitaxy, sestoft2018engineering}, polycrystalline Sn\cite{PendharkarTinArxiv} and amorphous Ta\cite{CarradArXiv191100460Cond-Mat2019}, albeit here with a much larger $\Delta$. Gate voltage-dependent bias spectroscopy (Supplementary Section 4.1) shows that the gap is highly robust in $V_\mathrm{G}$, excluding the possibility that the features attributed to $\Delta$ in Fig.~3 are due to e.g. Coulomb blockade.

The evolution of the superconducting gap in a parallel magnetic field, $B_{\parallel}$, is shown in Fig.~\ref{trans1}c. The gap remains hard over the entire 8.5~T range accessible in our cryostat, highlighted by the blue trace in Fig.~3b. The above-gap resonances likely arise from device-specific states in the tunnel barrier. Figure \ref{trans1}d shows the temperature-dependence of the superconducting gap. As temperature, $T$, is increased, $\Delta$ decreases and the edges broaden, up to the critical temperature $T_\mathrm{C} \sim 7$~K. Notably, the gap remains hard at $T=1.5$~K, and is still prominent at $T=4.5$~K. The values for $\Delta$, $T_\mathrm{C}$ and $B_\mathrm{C}$ of the hybrid InAs/Pb nanowires are each more than four times larger than for state-of-the-art Al-based hybrid devices\cite{chang2015hard, CarradArXiv191100460Cond-Mat2019, sestoft2018engineering, gazibegovic2017epitaxy, ZhangNature18}, and exceed the metrics reported for InSb/Sn hybrids\cite{PendharkarTinArxiv}. The dramatic increase in parameter space in terms of $\Delta$, $T_\mathrm{C}$ and $B_\mathrm{C}$ opens the possibility of performing semiconductor-superconductor hybrid measurements at kelvin temperatures and in entirely new transport regimes\cite{AmetScience2016}.

\section*{Coulomb blockade spectroscopy of InAs/Pb islands}

Essential requirements for semiconductor-superconductor heterostructures in the context of topological qubit development are for isolated hybrid segments -- `islands' -- to host quantised charge in units of $2e$, and the observation of single electron transport through bound states in finite $B_\parallel$ \cite{albrecht2016exponential, shen2018parity, Vaitiekenasfullshell18, PendharkarTinArxiv, sestoft2018engineering, vanHeckPRBmodel,AasenPRX16, shen2018parity}. Figure~4a shows an InAs/Pb island device, with the Pb selectively removed at either end and contacted by Ti/Au. The side gates control the island-contact coupling and the middle gate adds charge in discrete units. Spectroscopic operation is demonstrated in Fig.~4b, with $g$ plotted versus $V_\mathrm{SD}$ and $V_\mathrm{G}$, revealing periodic Coulomb diamonds due to the sequential addition of charge in units of $2e$ with an addition energy of $\sim 1.8$~meV, which corresponds to $8E_\mathrm{C}=(2e)^2/C$ \cite{shen2018parity}. The regions of negative differential conductance below $\Delta$ indicate the onset of long-lifetime quasiparticle tunneling\cite{albrecht2016exponential, shen2018parity, Vaitiekenasfullshell18}. For $|V_\mathrm{SD}|>1.2$~mV, the Coulomb diamond periodicity doubles in $V_\mathrm{G}$, since above-gap quasiparticle states support $1e$-periodic transport. 

The 2$e$-periodicity for $e|V_\mathrm{SD}|<\Delta$ in hard gap hybrids arises due to the odd charge states being lifted in energy by $\Delta$ compared to even charge states (Fig.~4c, top panel inset). For soft gap hybrids\cite{TakeiPRL2013, LeePRLsoftgap} the finite density of states within the gap may enable single electron tunneling and $1e$-periodicity at zero bias. The importance is that proposed topological qubit architectures rely on magnetic field induced zero energy bound states\cite{AasenPRX16}, and the topologically-protected qubit operation potentially offered by these devices is obviated if soft-gap states are present. Therefore, observing both zero-field $2e$-periodic transport and bound-state-induced $1e$-periodic transport are highly important steps in hybrid material development. 

\begin{figure}
  \centering
    \includegraphics[scale=1.05]{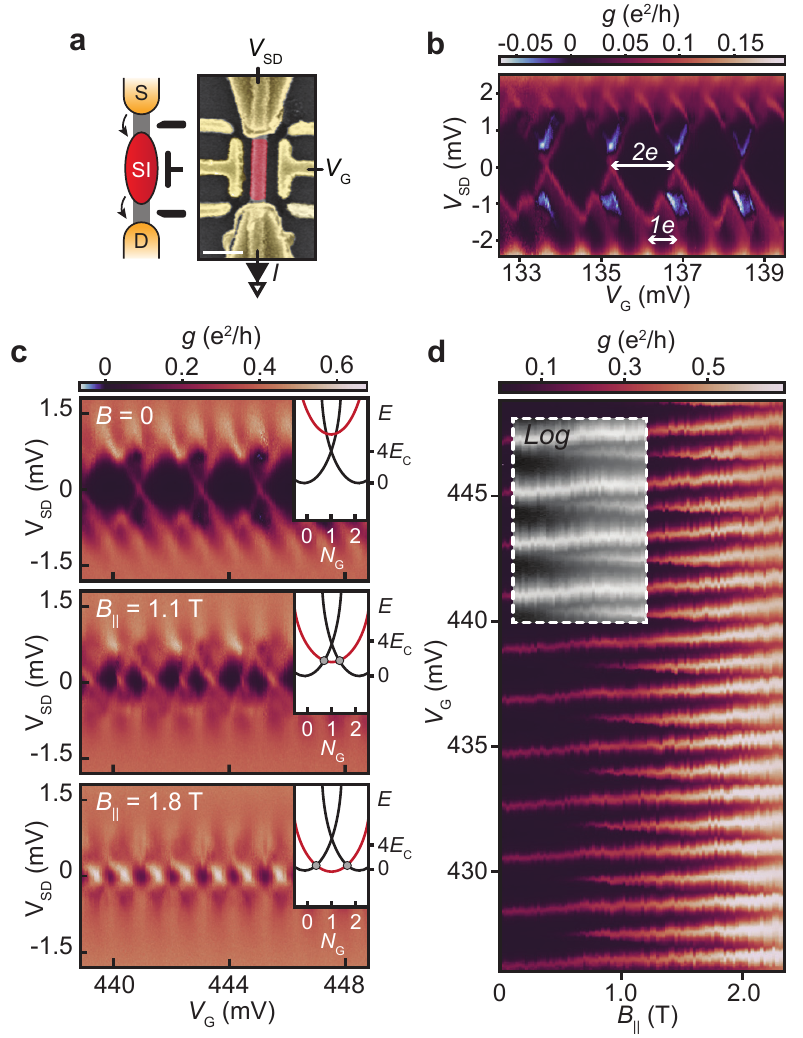}
    \caption{\textbf{Hybrid InAs/Pb island.} \textbf{a,} Schematic of a hybrid island device. Source, drain and superconducting island are denoted S, D and SI, respectively, with the black electrostatic gates used to control single electron (or Cooper pair) tunnelling events, illustrated by the arrows. False coloured electron micrograph of a InAs/Pb island device featuring Ti/Au contacts and gates (yellow), etched Pb segment (red) and narrow bare InAs segments (grey). Scale bar is 500 nm. \textbf{b,} $g$ vs $V_\mathrm{G}$ and $V_\mathrm{SD}$ showing Coulomb diamonds with spacing proportional to $2e$ ($1e$) for $|V_\mathrm{SD}|$ less than (greater than) $\Delta/e\ \sim\ 1.2$~mV, indicative of Cooper pair (single electron) charging. \textbf{c,} $g$ vs $V_\mathrm{SD}$ and $V_\mathrm{G}$ for $B_{\parallel}$ = 0, 1.1, and 1.8~T. Insets: Island energy, $E$, as a function of normalised gate voltage, $N_\mathrm{G}$. Black (red) parabolas indicate even (odd) ground state energies, $E_\mathrm{even}$ and $E_\mathrm{odd}$. Top panel: Cooper pair charging energy, $4E_\mathrm{C}<E_\mathrm{odd}$. Middle panel: $0<E_\mathrm{odd}<4E_\mathrm{C}$. Bottom panel: $E_\mathrm{even}=E_\mathrm{odd}=0$. \textbf{d,} Zero-bias conductance as a function of $V_\mathrm{G}$ and parallel magnetic field. Coulomb resonances split from a regular 2$e$ spacing ($4E_\mathrm{C}<E_\mathrm{odd}$), to even/odd ($0<E_\mathrm{odd}<4E_\mathrm{C}$) to a regular 1$e$ spacing ($E_\mathrm{even}=E_\mathrm{odd}=0$) as a function of parallel magnetic field. Inset: Conductance plotted on logarithmic scale (log$_{10}(g)$) to increase visibility of the low magnetic field behavior.}
    \label{island}
  \hfill
\end{figure}

The magnetic field dependent behaviour of our InAs/Pb island was studied using a gate voltage configuration giving a smaller addition energy of $\sim 1.2$~meV (Fig.~4c). At $B_\parallel=0$ (top panel), $2e$-periodic charging is observed below $\Delta$. Increased $B_\parallel=1.1$~T (Fig.~4c middle panel) causes a bound state to move towards zero energy\cite{deng2016majorana, albrecht2016exponential, shen2018parity, Vaitiekenasfullshell18}, and fall below $E_\mathrm{C}$. Odd-state occupation is now enabled (Fig.~4c middle panel inset), with the differing even/odd charging energies producing parity-dependent, even/odd spacing of Coulomb diamonds. Further increasing $B_\parallel=1.8$~T (lower panel) brings the bound state close to zero energy. 

Figure~4d presents the $B_\parallel$-dependence of $g$ versus $V_\mathrm{G}$ at zero bias, 
corresponding to the data in Fig.~4c. The $2e$-periodic peaks split with increasing $B_\parallel$ --- highlighted in the inset by the logarithmic greyscale plot --- with varying even/odd spacing (see Supplementary Fig.~13). The peak intensity asymmetry between even and odd peaks can be attributed to a difference in the electron-like and hole-like components of the bound states \cite{hansen2018probing,shen2018parity,PendharkarTinArxiv}. These data indicate the emergence of a zero energy hybrid bound state in the InAs/Pb island, consistent with the charging model presented in the insets of Fig.~4c, and previous experiments using Al- and Sn-based hybrids\cite{albrecht2016exponential, Vaitiekenasfullshell18, VaitiekenasPRL18sag, shen2018parity, CarradArXiv191100460Cond-Mat2019, PendharkarTinArxiv}. This demonstrates that InAs/Pb hybrids meet the criteria necessary for sustaining topological superconductivity. Future work including e.g. island-length-dependent studies\cite{albrecht2016exponential, Vaitiekenasfullshell18} or demonstrations of braiding\cite{AasenPRX16} will contribute to increased understanding towards this goal.

\section*{Discussion}
The formation of a flat, epitaxially matched single crystal along the entire nanowire confirms the expectations from theoretical considerations of thermodynamic driving forces and structural stability of Pb on InAs. The framework used to investigate elemental superconductor growth on semiconductor nanowires can be extended to other metals and semiconductors, providing increased scope for development of metal/semiconductor epitaxy. In addition to the structural perfection, the epitaxial InAs/Pb hybrids provide a large and hard induced superconducting gap, with corresponding high $T_\mathrm{C}$ and $B_\mathrm{C}$, offering a greatly extended experimental parameter space compared to current state-of-the-art materials.\cite{lutchyn2018majorana,CarradArXiv191100460Cond-Mat2019,GuskinNanoscale17,chang2015hard,gazibegovic2017epitaxy,ZhangNature18,PendharkarTinArxiv}

The observation of $2e$-periodicity and bound-state-induced $1e$-periodic transport in an InAs/Pb island represents an important step in the development of large $\Delta$, high $B_\mathrm{C}$ topological qubits\cite{PendharkarTinArxiv}. Moreover, implementing Pb in future devices may benefit from the strong intrinsic spin-orbit coupling of this heavy element superconductor, which has been used to induce spin-orbit coupling by proximity in, e.g., graphene\cite{klimovskikh2017spin,calleja2015spatial} and ferromagnets\cite{ruby2017exploring,nadj2014observation}. In the regime of stronger spin-orbit interaction one could potentially avoid the negative aspects of the predicted metallisation effect in Al based devices\cite{ReegPRB2018}, where the superconductor-semiconductor coupling increases the induced gap, while decreasing the desired spin-orbit interaction. Finally, recent scanning probe experiments on Pb-based topological systems\cite{nadj2014observation, menard2019isolated} can be complemented by exploring the strongly-coupled, two-band\cite{RubyPRL2015} superconducting nature of Pb in novel hybrid devices.

\section*{Methods}
The InAs nanowires were grown in a two step growth-process in a solid-source molecular beam epitaxy system. The nanowires were catalysed from electron-beam-lithography-defined Au particles and grown via the vapour-liquid-solid method. In the first step, InAs was grown axially for 30 minutes along the (111)B direction using As\textsubscript{4} and a substrate thermocouple temperature ($T_\mathrm{sub}$) of $447^\circ$C, resulting in nanowire lengths $\sim 7~\mu$m. Growth was then interrupted to increase the As cracker temperature from $500^\circ$C to $800^\circ$C, yielding As\textsubscript{2}, and the substrate temperature was lowered to ($T_\mathrm{sub}$) to $350^\circ$C over 20 minutes. The second growth step proceeded under these conditions, radially overgrowing the nanowires for 30 minutes.

After nanowire growth, the substrate was transferred to an attached chamber with a precooled stage without breaking ultra high vacuum (pressures $\sim10^{-10}$ Torr). The substrate was mounted on the cold substrate stage for 3 hours before any subsequent metal deposition to ensure a stable temperature ($T_\mathrm{sub}$) of $\sim$ 120 K. We note that less time on the cold substrate holder is preferred due to the possible unwanted deposition of foreign materials from the system. To enforce in-situ nanowire masking and only two facet depositions, as shown in Figs~2a,c,e, the wafer was aligned relative to the pre-defined nanowire array. Pb was deposited with a shallow angle and rate (3~Ås$^{-1}$) using electron beam evaporation. After the metal deposition the substrate was kept cold on the substrate holder for more than 10 hours to limit dewetting, and produced films with flatter morphology. 

SEM micrographs were produced by combining signals from the surface sensitive secondary electron detector and the atomic weight sensitive backscatter electron detector since Pb forms a flat film which is only seen clearly when using an atomic weight sensitive technique. STEM images were acquired using an FEI Titan 80-300 TEM equipped with a probe Cs corrector and operated at 300 kV. HAADF STEM images were taken with a beam convergence semi angle of $\sim 17.5$~mrad. The inner and outer collection semi angle of the HAADF detector was set to 54 and 270 mrad, respectively. Cross section specimens of the nanowires were prepared using an FEI Versa 3D focused ion beam scanning electron microscope (FIB SEM), following Pt deposition to protect the nanowires. A final polish of the TEM specimens in FIB SEM with a 2 kV and 27 pA ion beam was used to minimize the damage caused by the Ga ion beam. The micrograph in Fig.~1d was produced by drift correcting and summing the amplitude of 100 HR HAADF STEM micrographs with a short acquisition time. The STEM micrographs in  Figs~1e,f were noise reduced by removing the Fourier spectrum background amplitude. The structural simulation  of the hetero-epitaxial interface in Fig.~1 was performed using the software program Vesta\cite{momma2011vesta}. 

Devices (Figs~3 and 4) were fabricated using electron beam lithography and lift off techniques, with Ti/Au (5 nm/250 nm) thin films for contacts and gates. H$_2$O (MilliQ with resistivity $\rho>15$~M$\Omega$) was used as a selective etchant to remove Pb from the semiconductor surface, and Ar$^+$ milling used prior to contact deposition to remove native InAs oxides. Full details are given in the Supplementary Information. Standard voltage-biased lock-in techniques were used to measure the conductance of the devices in a dilution refrigerator with a base temperature of 25~mK. Prior to collecting the presented data on island devices (Fig.~4), the cross-coupling of the gates was measured. To collect the presented data, the gates were swept simultaneously according to the measured proportionality factors to ensure that $V_\mathrm{G}$ only altered island occupation and did not appreciably alter the island-lead coupling, or the chemical potential of the InAs segments.

\subsection*{Acknowledgements}
This work was funded by the Danish National Research Foundation, European Union’s Horizon 2020 research and innovation programme under grant agreement No.823717 (ESTEEM3), FETOpen grant no. 828948 (AndQC) and QuantERA project no. 127900 (SuperTOP), Villum Foundation project no. 25310, Innovation Fund Denmark's Quantum Innovation Center Qubiz, and the Carlsberg Foundation. J.d.B. acknowledges support by the Netherlands Organisation for Scientific Research (NWO/OCW), as part of the Frontiers of Nanoscience program. We thank M. Bjergfelt, A. Geresdi, T.S. Jespersen, J. Paaske, J.C.E. Saldaña and S. Vaitiekenas for useful discussions. C.B. S\o rensen is gratefully acknowledged for technical assistance and support. 

\subsection*{Competing interests}
The authors declare no competing interests

\subsection*{Supplementary Information} Supplementary information containing extended information on metal-on-semiconductor growth, structural characterisation and transport measurements is available at \url{https://sid.erda.dk/sharelink/eeL2CxBB6p}.

\end{document}